\documentclass{aa}
\usepackage[varg]{txfonts}
\usepackage{psfig}

\newcommand{\pedix}[2]{\ensuremath{#1_{\,\mbox{\scriptsize #2}}}}
\newcommand{\apix}[2]{\ensuremath{#1^{\,\mbox{\scriptsize #2}}}}
\newcommand{\pedap}[3]{\ensuremath{#1_{\,\mbox{\scriptsize #2}}^{\,\mbox{\scriptsize #3}}}}
\newcommand{\pedSM}[2]{\ensuremath{#1_{\,\mbox{\tiny #2}}}}
\newcommand{\errUD}[2]{\ensuremath{^{+#1}_{-#2}}}
\newcommand{\feka}{\ensuremath{\mbox{Fe~K}\alpha}}
\newcommand{\fekb}{\ensuremath{\mbox{Fe~K}\beta}}
\newcommand{\nika}{\ensuremath{\mbox{Ni~K}\alpha}}
\newcommand{\oiii}{\ensuremath{\mbox{[\ion{O}{iii}]}}}
\newcommand{\nev}{\ensuremath{\mbox{[\ion{Ne}{v}]}}}
\newcommand{\fermi}{\emph{Fermi}}
\newcommand{\galex}{{\sc GALEX}}
\newcommand{\herschel}{{\sc HERSCHEL}}
\newcommand{\rosat}{{\emph{ROSAT}}}
\newcommand{\spitzer}{{\emph{Spitzer}}}
\newcommand{\suzaku}{\emph{Suzaku}}
\newcommand{\swift}{{\emph{Swift}}}
\newcommand{\xmm}{{XMM-\emph{Newton}}}
\newcommand{\lum}{\ensuremath{\mbox{ergs~s}^{-1}}}
\newcommand{\flux}{\ensuremath{\mbox{ergs~cm}^{-2}\mbox{~s}^{-1}}}
\newcommand{\ionpar}{\ensuremath{\mbox{ergs~cm~s}^{-1}}}
\newcommand{\kms}{\ensuremath{\mbox{km~s}^{-1}}}
\newcommand{\nh}{\ensuremath{\mbox{cm}^{-2}}}
\newcommand{\nhsym}{\ensuremath{N_{\mbox{\scriptsize H}}}}
\newcommand{\arcdeg}{\ensuremath{^{\circ}}}
\newcommand{\kev}{\ensuremath{\,\mbox{\scriptsize keV}}}
\newcommand{\sfr}{\ensuremath{M_{\odot}\:\mbox{yr}^{-1}}}
\newcommand{\sfrsym}{\ensuremath{\mbox{\it SFR}}}
\newcommand{\normPL}{\ensuremath{\mbox{photons~keV}^{-1}\mbox{~cm}^{-2}\mbox{~s}^{-1}}}
\newcommand{\normGAUSS}{\ensuremath{\mbox{photons~cm}^{-2}\mbox{~s}^{-1}}}
\newcommand{\chidof}{\ensuremath{\chi^2/\mbox{d.o.f.}}}
\newcommand{\dchidof}{\ensuremath{\Delta\chi^2/\Delta\mbox{d.o.f.}}}
\newcommand{\src}{\object{1SXPS~J050819.8+172149}}

\title{AGN feedback in action: a new powerful wind in \src?
\thanks{Based on observations obtained with the \swift\ satellite.}}
\titlerunning{AGN feedback in action: a new powerful wind in \src?}
\author{
  L.~Ballo\inst{\ref{inst1}}\thanks{E-mail: lucia.ballo@brera.inaf.it (LB)}\and
  P.~Severgnini\inst{\ref{inst1}}\and
  V.~Braito\inst{\ref{inst2},\ref{inst3}}\and
  S.~Campana\inst{\ref{inst2}}\and
  R.~Della~Ceca\inst{\ref{inst1}}\and
  A.~Moretti\inst{\ref{inst1}}\and
  C.~Vignali\inst{\ref{inst4},\ref{inst5}}
}
\institute{Osservatorio Astronomico di Brera (INAF), via Brera 28, I-20121, Milano (Italy)\label{inst1}\and
Osservatorio Astronomico di Brera (INAF), via E. Bianchi 46, I-23807 Merate, LC (Italy)\label{inst2}\and
Department of Physics, University of Maryland, Baltimore County, Baltimore, MD 21250 (USA)\label{inst3}\and
Dipartimento di Fisica e Astronomia, Universit\`a degli Studi di Bologna, viale Berti Pichat 6/2, I-40127, Bologna (Italy)\label{inst4}\and
Osservatorio Astronomico di Bologna (INAF), Via Ranzani 1, I-40127, Bologna (Italy)\label{inst5}
}
\authorrunning{L.~Ballo et al.}

\begin{document}

\date{Received 21 May 2015 / Accepted 29 June 2015}

\abstract
{Galaxy merging is widely accepted to be a key driving factor in galaxy formation and evolution, while the feedback from actively accreting nuclei is thought to regulate the black hole-bulge coevolution
and the star formation process.
}
{In this context, we focused on \src, a local ($z=0.0175$) Seyfert~1.9 galaxy ($\pedix{L}{bol}\sim 4\times10^{43}\,$\lum).
The source belongs to an infrared-luminous interacting pair of galaxies, characterized by a luminosity for the whole system (due to the combination of star formation and accretion)
of $\log (\pedix{L}{IR}/\pedix{L}{\sun})=11.2$.
We present here the first detailed description of the $0.3-10\,$keV spectrum of \src, monitored by \swift\ with $9$ pointings performed in less than $1\,$month.
}
{The X-ray emission of \src\ is analysed by combining all the \swift\ pointings, for a total of $\sim 72\,$ks XRT net exposure.
The averaged \swift-BAT spectrum from the 70-month survey is also analysed.
}
{The slope of the continuum is $\Gamma\sim 1.8$, with an intrinsic column density of $\sim 2.4\times10^{22}\,$\nh, and a de-absorbed luminosity of $\sim4\times10^{42}\,$\lum\
in the $2-10\,$keV band.
Our observations provide a tentative ($2.1\sigma$) detection of a blue-shifted \ion{Fe}{xxvi} absorption line (rest-frame $E\sim 7.8\,$keV), thus suggesting  
the discovery for a new candidate powerful wind in \src.
The physical properties of the outflow cannot be firmly assessed, due to the low statistics of the spectrum and to the observed energy of the line, 
too close to the higher boundary of the \swift-XRT bandpass.
However, our analysis suggests that, if the detection is confirmed, the line could be associated with
a high-velocity ($\pedix{v}{out}\sim 0.1c$) outflow most likely launched within $80\,\pedix{r}{S}$.
To our knowledge this is the first detection of a previously unknown ultrafast wind with \swift.
The high column density 
suggested by the observed equivalent width of the line (EW$\sim -230\,$eV, although with large uncertainties), 
would imply a kinetic output strong enough to be comparable to the AGN bolometric luminosity.
}
{}


\maketitle

\section{Introduction}\label{sect:intro}

The observational evidence for the presence of inactive Super Massive Black Holes (SMBHs; $\pedix{M}{BH}\sim10^{6}-10^{10}\,\pedix{M}{\sun}$) at the centre of most, if not all, the local galaxies, 
and the observed correlation between several properties of the galaxy's bulge and the central SMBH mass \citep{ferrarese00,
gebhardt00}, suggest that the SMBH accretion and the assembly of the galaxies bulges are intimately related 
\citep[see][for a recent review]{kormendy13}.
Funnelling of gas in the nuclear regions, as triggered by galaxy interactions, can activate both efficient accretion onto the SMBH, and a burst of star formation.
A key ingredient in regulating their evolution should be the feedback from the 
Active Galactic Nuclei (AGN);
being conservative, while building its mass the SMBH can release an amount of energy larger than $\sim 30$ times the binding energy of the host bulge \citep[see a review in][]{fabian12}.
Even if only a small fraction of this energy is transferred to the gas in the galaxy, then an active nucleus can have a profound effect on the evolution of its host 
\citep{dimatteo05}.
Powerful (kinetical or radiative) outflows of gas driven by luminous quasars are invoked as a 
key mechanism to blow away the gas in the galaxy and thereby quench star formation, coincidentally starving the SMBH of fuel \citep{king15}.

AGN winds with a range of physical properties have been revealed by observations at various energies, from radio up to X-rays.
Outflows of molecular or neutral atomic gas, with velocities up to $\sim 1000-2000\,$\kms\ 
and extending on kpc scales, have been observed at mm \citep[e.g.,][]{feruglio10,cicone14}
and radio frequencies \citep[e.g.,][]{morganti05,teng13}
in a few dozen AGN in dusty star forming sources and/or radio galaxies.
Mass outflows of ionized gas with similar velocities at distances consistent with the narrow line region zone have been detected in the optical/ultraviolet (UV),
both in the \oiii\ emission line profiles \citep[e.g.,][]{crenshaw05,canodiaz12,cresci15}, and through the observation of broad absorption line systems \citep[e.g.,][]{dai08,borguet13}. 
In X-rays, mildly ionized warm absorbers are observed in more than half of unobscured AGN \citep{crenshaw03,blustin05}.
The observed velocities of $\sim 500-1000\,$\kms\ imply a kinetic power rather low when compared to the bolometric luminosity.
However, \citet{crenshaw12} found that, 
when summed over all the absorbers, the total power carried by these structures can reach, in some cases, the minimum level required 
for AGN feedback 
\citep[$L/\pedix{L}{bol}\sim0.5-5$\%; e.g.,][]{dimatteo05,hopkins10}.
In the last years, highly blue-shifted Fe K-shell absorption lines at rest-frame energies $E \gtrsim 7\,$keV, observed in \xmm\ or \suzaku\ spectra of luminous AGN, revealed the presence
of high column density ($\nhsym\sim10^{23}\,$\nh) and fast ($v > 0.1c$) winds \citep[e.g.,][]{tombesi10,gofford13,tombesi15}. 
Their derived kinetic power is systematically higher than the minimum fraction of bolometric luminosity required by AGN feedback models in order to regulate the growth of an SMBH and its galactic bulge
\citep[e.g.,][]{hopkins10}.
Only very recently, outflows over a range of scales have been detected and studied within the same source, thus allowing us to explore
the connection between large-scale molecular outflows and accretion-disk activity \citep{feruglio15,tombesi15}.

\vspace{0.2 truecm}
Finding observational evidence of the effects of such 
accretion-related 
feedback and characterising the magnitude of mass outflows from AGN
are among the major challenges of the current extragalactic astronomy.
Here we present $9$ coadded \swift\ observations (performed in less than $1\,$month) of the interacting infrared (IR) galaxy \src.
The spectrum, the first for this source covering the energy range $E\sim7-10\,$keV,
provides us with a tentative detection of a possible new ultrafast wind ($\pedix{v}{out}\sim 0.1c$).
To our knowledge, this is the first time that a previously unknown outflow is revealed by \swift.

The source is described in Sect.~\ref{sect:src}, while the analysis of the \swift-XRT data is presented in Sect.~\ref{sect:anal} and the results are summarised in Sect.~\ref{sect:summ}.
Throughout the paper we assume a flat $\Lambda$CDM cosmology with $\pedix{H}{0}=71\,$\kms~Mpc$^{-1}$, 
$\Omega_\Lambda=0.7$ and $\pedix{\Omega}{M}=0.3$.


\section{\src}\label{sect:src}

The position of the \swift\ source \src\ \citep{evans14} is coincident with the center of \object{CGCG 468-002~NED01}, a local ($z=0.0175$) galaxy belonging to the interacting pair \object{CGCG 468-002} 
(projected distance between the center of the galaxies of $\sim 29.4\arcsec$, corresponding to $\sim 10\,$kpc at the source redshift).

The high $8-1000\,\mu$m luminosity
of the system, $\log (\pedix{L}{IR}/\pedix{L}{\sun})=11.2$ \citep{armus09}, implies a classification as Luminous Infrared Galaxy 
\citep[LIRG, defined as having $\pedix{L}{IR}\simeq 10^{11-12}\,$\pedix{L}{\sun};][]{sanders96}.
Included for this reason in the Great Observatories All-Sky LIRG Survey \citep[GOALS;][]{armus09}, the system has been targeted with several observational facilities, 
and multiwavelength information (both spectroscopic and photometric) has been collected.
The available data span from the UV \citep[\galex;][]{howell10} up to the mid-IR 
\citep[\spitzer/IRS, IRAC, and MIPS;][see also \citealt{valiante09,aherrero12,aherrero13}] {diazsantos10,diazsantos11,petric11,inami13,stierwalt13,stierwalt14} and far-IR 
\citep[\herschel/PACS;][]{diazsantos13} bands.
Photometry from the 2MASS Redshift Survey is reported by \citet{huchra12}.
At radio wavelengths, the relatively strong ($\pedix{S}{$\nu$}\sim 34\,$mJy) NVSS emission \citep{condon98}, detected halfway between the nuclei and slightly elongated in their direction, is probably 
due to the combined contribution of both galaxies.

The system is an early-stage merger \citep{stierwalt13} known to host an AGN optically classified as Seyfert~1.9 \citep[see e.g.][]{motch98,vcv01,kollatschny08}.
Mainly based on the detection of the high-ionization line \nev\ in the \spitzer/IRS spectra, the Seyfert~1.9 nucleus has been associated with
the western galaxy of the system, coincident with \src\ 
\citep[see also the optical classification reported by \citealt{aherrero13}]{petric11,aherrero12,stierwalt13,stierwalt14}.
The \spitzer\ spectra are suggestive of a
relatively unobscured AGN (consistent with the optical classification as Seyfert~1.9), energetically important in the mid-IR \citep{stierwalt13}.
From a decomposition of the mid-IR spectra into AGN and starburst components, \citet{aherrero12} estimated a bolometric luminosity due to the accretion of $\pedix{L}{bol}\sim 10^{10}\pedix{L}{\sun}$.
Instead, no signature of active accretion is found for the eastern galaxy, which shows all the typical properties of a star forming source both in the mid-IR spectra and from the UV photometry.

The BH mass in \src\ is among the highest observed in local LIRGs, $\pedix{M}{BH}\sim 1.15\times 10^{8}\,\pedix{M}{\sun}$ 
\citep[as calculated from the velocity dispersion of the core of the \oiii$\lambda5007$ line;][]{aherrero13}.
This implies that the black hole is radiating at a low fraction of its Eddington luminosity\footnote{The Eddington luminosity,
defined as $\pedix{L}{Edd} = 4\pi G\pedix{M}{BH}\pedix{m}{p}c/\pedix{\sigma}{T}\simeq1.3 \times 10^{38} \pedix{M}{BH}/\pedix{M}{\sun}\,$[\lum], 
represents the exact balance between inward gravitational force and
outward radiation force acting on the gas, assumed to be of ionised hydrogen in a spherical configuration.}, with Eddington ratio
$\log\pedix{\lambda}{Edd}\equiv\log\pedix{L}{bol}/\pedix{L}{Edd}\sim -2.5$.
The star formation rate (\sfrsym) derived for this source falls in the lower tail of the distribution found for local LIRGs \citep[nuclear $\sfrsym\sim1-3\,$\sfr, 
integrated $\sfrsym\sim3-4\,$\sfr,][]{aherrero13}, implying a ratio between \sfrsym\ and BH accretion rate of $\log \sfrsym/\pedix{\dot{m}}{BH}\sim 2$ 
(by assuming a mass-energy conversion efficiency $\epsilon = 0.1$), similar to the values found for Seyfert galaxies.


\section{XRT data analysis}\label{sect:anal}

%
\begin{table}
\begin{minipage}[t]{0.5\textwidth}
 \caption{\swift-XRT monitoring observation log for \src. Net count rates are in the $0.3-10\,$keV energy range.}
 \label{tab:log}
 \centering
{
  \begin{tabular}{@{}l r@{\extracolsep{0.cm}-}c@{\extracolsep{0.cm}-}l@{\extracolsep{0.2cm}} c c@{}}
  \hline  \hline
Obs. ID & \multicolumn{3}{c}{Start date} & Net count rate  &  Net exp. time\\
        & \multicolumn{3}{c}{} & [$10^{-2}\,$counts~s$^{-1}$]        &      [s]
 \vspace{0.1cm} \\
 \hline
 \vspace{-0.2cm} \\
 00049706003 &  2014 & 10 & 15 & $7.5 \pm 0.4$ & $4235$ \\
 00049706004 &  2014 & 10 & 16 & $13.1\pm 0.3$ & $14740$ \\
 00049706005 &  2014 & 10 & 21 & $10.0\pm 0.7$ & $2308$ \\
 00049706006 &  2014 & 10 & 22 & $7.3 \pm 0.3$ & $7325$ \\
 00049706007 &  2014 & 10 & 23 & $7.5 \pm 0.3$ & $6533$ \\
 00049706008 &  2014 & 10 & 27 & $5.8 \pm 0.3$ & $6478$ \\
 00049706009 &  2014 & 10 & 28 & $5.9 \pm 0.3$ & $6218$ \\
 00049706010 &  2014 & 10 & 30 & $4.6 \pm 0.3$ & $6333$ \\ 
 00049706011 &  2014 & 11 & 11 & $7.6 \pm 0.2$ & $17950$
    \vspace{0.1cm} \\
   \hline
  \end{tabular}
 }
\end{minipage}
\end{table}

Besides the basic analysis reported in the \swift-XRT point source catalogue \citep{evans14},
in the soft-medium X-ray energy range ($E\lesssim 10\,$keV) the only published information up to now comes from the \rosat\ All Sky Survey \citep[see][and references therein]{kollatschny08}, 
providing a soft 
X-ray luminosity (de-absorbed by our Galaxy) of $\pedix{L}{0.1-2.4\kev}\sim6.6\times10^{41}\,$\lum.
At higher energies, \src\ has been detected by the \swift-Burst Alert Telescope \citep[BAT; see][]{batcat70m}, 
while \citet{ackermann12} reported only a \fermi\ $95$\% confidence level upper limit 
of $\pedix{L}{0.1-10\,\mbox{GeV}}\sim1.6\times10^{42}\,$\lum.

Recently, our group has been awarded a \swift\ \citep{swift} program for this source (PI P.~Severgnini): $9$ pointings performed with the X-ray telescope \citep[XRT;][]{xrt} 
in the standard photon counting (PC) mode between 2014-10-15 and  2014-11-11, for a total of $\sim 72\,$ks net exposure (ObsID from 00049706003 to 00049706011; see Table~\ref{tab:log}).

We generated images, light curves, and spectra, including the background and ancillary response files, with the online XRT data product 
generator\footnote{http://www.swift.ac.uk/user\_objects} \citep{evans07,evans09}; the appropriate spectral response files have been identified in the calibration database.
The source appears point-like in the XRT image and centered at the position of the western nucleus, without any evident elongation toward the position of the second galaxy. 
We note that at the angular resolution of XRT, $18\arcsec$ half-power diameter, the emission of the two galaxies, located at a distance of $\sim 29.4\arcsec$, can be resolved;
by assuming a power-law model with $\Gamma=2$, we estimated a $3\sigma$ upper limit to the $0.3-10\,$keV emission at the eastern source position of $\sim 7\times 10^{-14}\,$\flux.

Source events were extracted from a circular region with a radius of $20\,$pixels 
\citep[which corresponds to an Encircled Energy Fraction of 90\%,][$1\,$pixel $\sim2.36\,$arcsec]{moretti05}, while background events were
extracted from an annulus region centred on the source with inner and outer radii of $60$ and $180\,$pixels, respectively; 
all sources identified in the image
were removed from the background region.

We extracted a light curve binned at the duration of each individual observation.
\src\ was detected in all observations, with $0.3-10\,$keV signal-to-noise ratios ($S/N$) ranging from $15$ to $44$.
The average count rates in the total ($0.3-10\,$keV), soft ($0.3-2\,$keV), and hard ($2-10\,$keV) XRT energy ranges are $\sim 0.09$, $\sim 0.03$, and $\sim 0.07\,$counts~s$^{-1}$, respectively.
Small deviations from these values, of a factor lower than $2$, are observed in the light curves;
however, there is no evidence of spectral variability in the ratio of count rates observed in the hard and soft bands.
We do not find any significant pile-up problem.

In order to increase the statistics, we co-added the XRT datasets.
Source and background spectra were extracted from the merged event lists, and the former was binned in order to have at least $20$ total counts per energy channel. 
The net count rates in the $0.3-10\,$keV, $0.3-2\,$keV, and $2-10\,$keV energy ranges are $(8.2\pm0.1)\times10^{-2}$, $(2.3\pm0.1)\times10^{-2}$, and 
$(5.8\pm0.1)\times10^{-2}\,$counts~s$^{-1}$, respectively.
The $S/N$ achieved in the same energy ranges are $76$, $40$, and $64$, respectively.

Spectral fits 
were performed 
in the $0.3-10\,$keV energy range using the X-ray spectral fitting package XSPEC \citep{xspec} v12.8.2.
Uncertainties are quoted at the $90$\% confidence level for one parameter of interest ($\Delta\chi^2 = 2.71$). 
All the models discussed in the following assume Galactic absorption with a column density of $\pedix{N}{H,Gal}=1.84\times 10^{21}\,$\nh\ \citep{nh}.
To model both Galactic and intrinsic absorptions we used the {\sc (z)phabs} model in XSPEC, adopting cross-sections and abundances of \citet{wilms}.

\begin{figure}
 \centering
 \resizebox{1.\hsize}{!}{\includegraphics[angle=270]{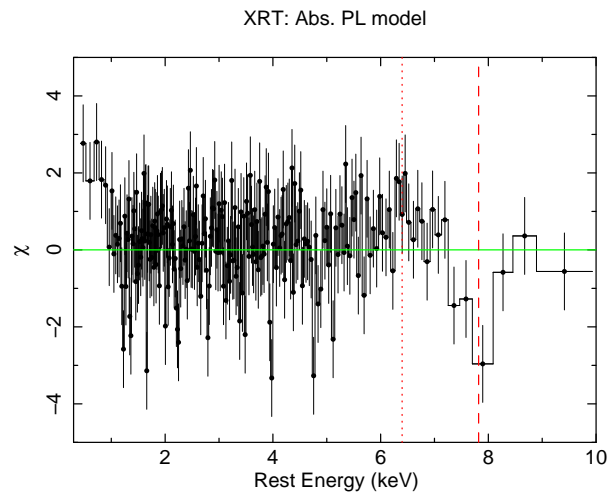}}
 \caption{
 Residuals, plotted in terms of sigmas with error bars of size one,
 for the XRT spectrum, modelled with an absorbed power-law component ($\Gamma\sim 1.5$,  $\nhsym\sim 1.8\times 10^{22}\,$\nh).
A deep absorption trough is seen in the iron K band, possibly associated with a broad absorption feature (dashed red line).
The dotted red line marks the position where the neutral \feka\ emission line is expected (and marginally detected).
}
\label{fig:abpl_ra}%
\end{figure}
%

%
\begin{table*}
\begin{minipage}[t]{1\textwidth}
 \caption{Summary of the \swift\ (XRT$+$BAT) parameters of the best-fit model described in Sect.~\ref{sect:anal}, where a high-energy absorption feature is 
 superimposed to a continuum composed by an absorbed power law plus a reflection component, with the addition of a scattered power law. }
 \label{tab:res}
 \centering
\resizebox{1\textwidth}{1cm} 
{
 \footnotesize
  \begin{tabular}{c@{\extracolsep{0.2cm}} c@{\extracolsep{0.2cm}} c@{\extracolsep{0.2cm}} c@{\extracolsep{0.2cm}} c@{\extracolsep{0.2cm}} c@{\extracolsep{0.2cm}} c@{\extracolsep{0.2cm}} c@{\extracolsep{0.2cm}} c@{\extracolsep{0.2cm}} c@{\extracolsep{0.2cm}} c@{\extracolsep{0.2cm}} c@{\extracolsep{0.2cm}}}
   \hline\hline       
      \multicolumn{5}{c}{Direct, reflected and scattered continuum} & \multicolumn{4}{c}{Absorption line} &  & &  \\
   \cline{1-5}\cline{6-9}
   \nhsym & $\Gamma$ & \pedSM{A}{pl} & $R$ & \% scatt. & $E$ & \pedSM{\sigma}{E} & \pedSM{A}{E} & EW & \pedSM{\log F}{2-10\kev} & \pedSM{\log L}{2-10\kev} & \chidof \\
  (1) & (2) & (3) & (4) & (5) & (6) & (7) & (8) & (9) & (10) & (11) & (12)  
    \vspace{0.1cm} \\
   \hline
    \vspace{-0.2cm} \\
    $2.4 \pm 0.3$ & $1.8 \pm 0.1$ & $1.8\errUD{0.4}{0.3}\times 10^{-3}$ & $1.4 \pm 0.4$ & $2.1 \pm 0.8$ & $7.8\pm 0.3$ & $<270$ & $-(1.3\errUD{1.3}{0.9})\times 10^{-5}$ & $-(230\errUD{390}{170})$ & $-11.19\pm 0.22$ & $42.61\pm 0.18$ & $241.5/228$ 
    \vspace{0.1cm} \\
   \hline
  \end{tabular}
 }
 \tablefoot{\footnotesize Errors are quoted at the $90$\% confidence level for $1$ parameter of interest ($\Delta\chi^2=2.71$).
 \footnotesize Col. (1): Intrinsic column density, in units of $10^{22}\,$\nh.
 \footnotesize Col. (2): Power-laws (absorbed and unabsorbed) and reflection component photon index.
 \footnotesize Col. (3): Absorbed power-law normalisation, in units of \normPL.
 \footnotesize Col. (4): Reflection fraction.
 \footnotesize Col. (5): Percentage of scattering fraction, defined as the ratio of the unabsorbed and absorbed power-law normalizations. 
 \footnotesize Col. (6): Rest-frame energy centroid of the Gaussian absorption line, in units of keV.
 \footnotesize Col. (7): Absorption line width, in units of eV.
 \footnotesize Col. (8): Gaussian absorption line normalisation, in units of \normGAUSS.
 \footnotesize Col. (9): Absorption line equivalent width, in units of eV.
 \footnotesize Col. (10): Observed flux (de-absorbed by our Galaxy) in the $2-10\,$keV energy band, in units of \flux.
 \footnotesize Col. (11): Intrinsic luminosity in the $2-10\,$keV energy band, in units of \lum.
 \footnotesize Col. (12): $\chi^2$ and number of degree of freedom.
 }
\end{minipage}
\end{table*}

A fit with a simple absorbed power law clearly provides a poor representation of the XRT data ($\chidof = 280.8/229$); the photon index is $\Gamma = 1.49 \pm 0.09$ 
and the column density is $\nhsym = (1.8\pm 0.2)\times 10^{22}\,$\nh.
Above $\sim 5\,$keV, i.e. in the energy range where the iron K complex is expected, residuals  are present both in emission and
in absorption (see Fig.~\ref{fig:abpl_ra}), while a big bump below $1\,$keV suggests the emerging of an additional component in the soft band.
Assuming that the neutral absorber only partial covers the central source ({\sc zpcfabs} model in XSPEC) still leaves residuals in the soft band
($\chidof = 271.0/228$); we found a covering fraction$\,\gtrsim 0.95$, with $\Gamma = 1.48 \pm 0.04$ and $\nhsym = (2.1\pm 0.1)\times 10^{22}\,$\nh.

In principle, the emission observed at low energies can be associated with the accreting nucleus (i.e., due to scattering off optically thin ionised gas) and/or the host galaxy 
(a soft thermal emission is a characteristic signature in all known starburst galaxies).
Phenomenologically, the observed soft excess can be accounted for equally well by adding to the previous model: {\it 1)} an unabsorbed power law with photon index tied to the primary one 
($\chidof = 258.4/228$); the best-fit parameters are: $\Gamma = 1.57 \pm 0.09$, $\nhsym = (2.2\pm 0.2)\times 10^{22}\,$\nh, and ratio of
the power-law normalizations $\pedix{A}{scatt}/\pedix{A}{intr}\sim3$\%; or {\it 2)} a thermal component 
({\sc mekal} model in XSPEC; $\chidof = 259.3/227$): in this case we found $\Gamma = 1.53 \pm 0.09$, $\nhsym = (1.9\pm 0.2)\times 10^{22}\,$\nh, and $kT=0.31\errUD{0.24}{0.09}\,$keV.
In the latter parametrisation, the luminosity attributed to the thermal component is $\pedap{L}{0.5-2\kev}{mek}\sim 2.6\times 10^{40}\,$\lum, which would imply a 
$\sfrsym\sim0.5-2.9\,$\sfr\ \citep[e.g.,][]{ranalli03,mashesse08,mineo12}, consistent with the nuclear star formation properties of the host (see Sect.~\ref{sect:src}).
In both cases, the luminosity observed in the range covered by \rosat, $\pedix{L}{0.1-2.4\kev}\sim 5.6\times 10^{41}\,$\lum, is in agreement
with the value derived by
\citet{kollatschny08}.

While the most plausible hypothesis is that both components contribute to the emission observed at low energies in the XRT spectrum, 
the quality of the present data does not allow us to 
discriminate between the two contributions, and even less to 
disentangle 
them.
In the following, we assume an unabsorbed power law, checking that the inclusion of a thermal component in place of the power law does not affect the main results presented here.

The addition of a narrow ($50\,$eV) Gaussian emission line to the absorbed plus unabsorbed power laws results in an improvement in the fit 
($\dchidof=11.3/2$).
The line parameters are $E = 6.40\pm0.09\,$keV, consistent with neutral \feka, $I = 1.5\pm0.7\times 10^{-5}\,$\normGAUSS, and EW$\sim 200\,$eV.
The strength of the \feka\ line, coupled with the hard photon index ($\Gamma\sim1.62$), could suggest the presence of neutral reflection \citep {reynolds94,matt96,matt00}.
This possible component was then included in the model by replacing the narrow Gaussian line with a {\sc pexmon} component \citep{pexmon}, an additive model self-consistently incorporating 
the Compton-reflected continuum from a neutral slab combined with emission from \feka, \fekb, \nika\ and the \feka\ Compton shoulder.

To improve the determination of the slope of the primary power law and the amount of reflection,
it is 
fundamental
to know the shape and intensity of the emission at energies higher than $10\,$keV.
Therefore we fitted the XRT data simultaneously with the averaged \swift-BAT spectrum of \src\ obtained from
the 70-month survey archive\footnote{http://swift.gsfc.nasa.gov/results/bs70mon/SWIFT\_J0508.1p1727} (SWIFT~J0508.1+1727).
The data reduction and extraction procedure of the eight-channel spectrum is described in \citet{batcat70m}. 
To fit the pre-processed, background-subtracted BAT spectrum, we used the latest calibration response
as of 2013 May.
\src\ was detected in the $14-100\,$keV band with a count rate of $(3.9 \pm 0.3) \times 10^{-5}\,$counts~s$^{-1}$, which corresponds to a $14-195\,$keV flux of 
$2.6\pm 0.5 \times 10^{-11}\,$\flux\ \citep{batcat70m}.

During the fit, the only free parameter of the {\sc pexmon} component was the reflection scaling factor, $R$.
We tied the {\sc pexmon} photon index and normalization to that of the primary power law, and we fixed the cutoff energy at $1000\,$keV (i.e., consistent with no
measurable cut-off), the inclination angle at $60\arcdeg$, and the abundances of heavy elements at their Solar values.
We also allow to vary the cross-normalisation factor between the XRT data and the average BAT spectrum.

The baseline model then consists of an absorbed power law and an unabsorbed one plus a reflection component: the photon index and intrinsic absorption of the primary emission are 
$\Gamma=1.75\pm0.09$  \citep[in agreement with the values typically observed in unobscured AGN;][]{piconcelli05,mateos10,corral11}
and $\nhsym=(2.4\pm0.2)\times 10^{22}\,$\nh, while the reflection fraction 
and the strength of the scattered component are $R\sim 1.1$ and $\pedix{A}{scatt}/\pedix{A}{intr}\sim2$\%.
The cross-normalisation factor between the XRT data and the average BAT spectrum
is $0.96$.

\begin{figure}
 \centering
 \resizebox{1.\hsize}{!}{\includegraphics[angle=270]{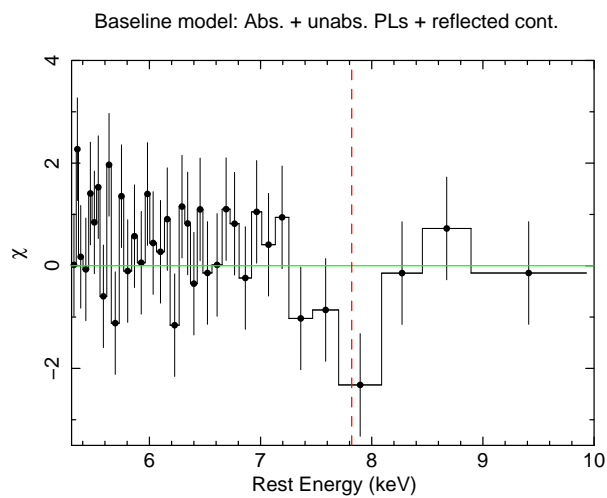}}
 \caption{
 Residuals, plotted in terms of sigmas with error bars of size one,
 for the XRT spectrum above $5.3\,$keV. The adopted model is an absorbed power law and a reflection component 
 ($\Gamma\sim 1.8$,  $\nhsym\sim 2.4\times 10^{22}\,$\nh, $R\sim1.1$).
 An unabsorbed power law with a $\sim 2$\% of intensity of the intrinsic one accounts for the emission below $\sim 1\,$keV
 (actually, below the reported energy interval).
 The centroid of the absorption line when one more Gaussian is included in the model is marked with a red vertical line (absorption line at $E\sim7.8\,$keV).}
 \label{fig:cont_ra}%
\end{figure}

\vspace{0.2 truecm}
Although this model provides an acceptable description of the broad-band ($0.3-100\,$keV) X-ray continuum of \src\ ($\chidof=250.2/231$),
it is not able to account for the residuals observed above $7\,$keV, 
confirming the presence of a deep absorption trough at about $8\,$keV
(see Fig.~\ref{fig:cont_ra}).
Considering only the $41$ XRT bins
between $5$ and $9\,$keV,
the fit statistics provided by the model is  $\chidof=44.2/36$.
These residuals can be accounted for by adding a Gaussian absorption line, with a rest-frame energy of $E=7.8\pm 0.3\,$keV and an equivalent width EW$=-(230\errUD{390}{170})\,$eV
($\dchidof=8.3/3$ in the $5-9\,$keV interval).
The line is marginally resolved; allowing its width to vary, we can set only an upper limit $\pedix{\sigma}{E}< 270\,$eV.

\begin{figure}
 \resizebox{1.\hsize}{!}{\includegraphics[angle=270]{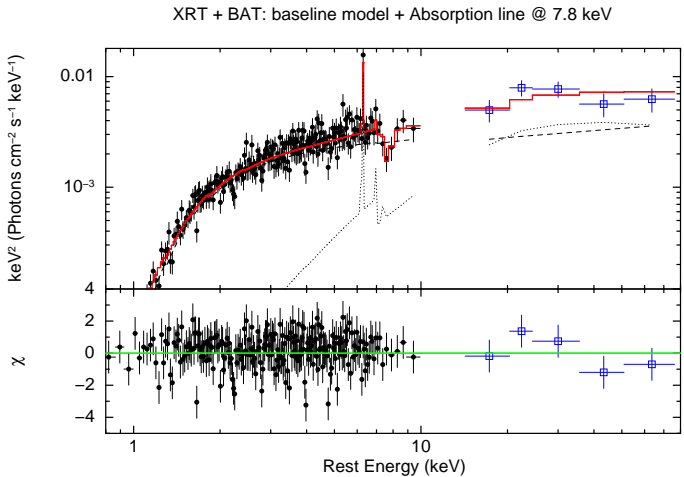}}
 \caption{ 
 {\it Upper panel:} Unfolded XRT data (in black) and average BAT spectrum (in blue) fitted including a Gaussian absorption line ($E\sim 7.8\,$keV, EW$\sim -230\,$eV and $\pedix{\sigma}{E}\sim 190\,$eV). 
 The continuum (red continuous line) is modelled with an absorbed plus an unabsorbed power law and a reflection component.
 {\it Lower panel:} Relevant residuals, plotted in terms of sigmas with error bars of size one.
}
\label{fig:bf}%
\end{figure}

The 
best-fitting 
parameters
are reported in Table~\ref{tab:res}, while the unfolded XRT and BAT spectra are shown in Fig.~\ref{fig:bf}.
We note that the significance of the detection can depend on the accuracy of the determination of the continuum shape.
In particular, we checked if the observed shape can be explained with a mix of a stronger reflection edge plus a steeper and lower continuum.
However, even if we assume the combination of reflection strength and intrinsic spectral shape that minimises the intensity of the trough, 
the normalisation of the Gaussian line is still inconsistent with the null value at a confidence level of $90$\%, in agreement
with the significance of the detection derived from the simulations (see below).

The high energy range where the line is observed, very near to the end of the XRT bandpass, could raise concerns of possible artefacts due to the background.
However, the $S/N$ and number of source counts collected between $8$ and $10\,$keV ($\sim 7$ and $\sim56\,$counts, respectively; last $3$ bins in Fig.~\ref{fig:cont_ra}) 
strongly support that the spectral rising that defines the line is real.
We performed extensive simulations testing the null hypothesis that the spectrum is well fitted by a model that does not include the $7.8\,$keV absorption feature, as done in 
\citet[see also \citealt{porquet04}]{markowitz06}. 
Briefly, 
to take into account the observed background and the uncertainty in the continuum, we first generated a fake spectrum for a $72\,$ks exposure assuming the best fit found for the continuum.
We then fitted
this model to the fake spectrum, and starting from the new best-fit parameters we re-run a simulation with the same exposure.
The baseline model has been fitted to this final fake spectrum, and the derived $\chi^2$ has been compared with the minimum value of $\chi^2$ obtained when 
a narrow ($\sigma=10\,$eV) Gaussian component was included.
We stepped the centroid energy of the absorption line over the $6.5-9\,$keV range in increments of $0.125\,$keV, fitting separately each time to derive the lowest value of $\chi^2$.
The whole process has been repeated $400$ times, and we estimated a $4$\% probability of detecting a similar feature by chance.

The observed energy of the line suggested by the data
is not consistent with any of the atomic transitions\footnote{http://physics.nist.gov/PhysRefData/ASD/lines\_form.html} expected at energies $\gtrsim7\,$keV \citep[e.g.,][]{kallman04}. 
The most likely explanation is that the centroid of the line is blueshifted, implying that the material responsible for the observed feature is outflowing.
Prime candidates for the origin of the line are the inner K-shell resonances from moderately-highly ionized \feka, as observed in other AGN.
Indeed, assuming K$\beta$ absorption by moderately ionized Fe, we would expect to observe also strong K$\alpha$ absorption by the same species
\citep[see e.g. the detailed discussion in][]{markowitz06}.
Being conservative, we can identify the line with the \ion{Fe}{xxvi} at $E = 6.966\,$keV resonant absorption; in this case, the observed centroid would indicate 
a substantial blueshifted velocity of $(0.11\pm 0.03)c$.
An origin in \feka\ at a lower ionization would imply an even higher blueshift.

The observed outflow velocity of $\pedix{v}{out}\sim 0.11c$ would translate in a lower limit on the radial distance, corresponding to the escape radius at which 
the material is able to leave the system.
When a spherical geometry is assumed, this limit is $R\geq\pedix{R}{esc} = 2GM/\pedap{v}{out}{2}$,
where $M$ is the enclosed mass producing the inward gravitational force.
Under the reasonable hypothesis that this corresponds to the mass of the central BH,
assuming the estimate reported in the literature ($\pedix{M}{BH}\sim 1.15\times 10^{8}\,\pedix{M}{\sun}$, see Sect.~\ref{sect:src}), we have
$\pedix{R}{esc}\sim 2.7\times 10^{15}\,$cm.
This translates to a possible launch radius of $10^{-3}\,$pc, or $\sim 80\,\pedix{r}{S}$ (Schwarzschild
radii, $\pedix{r}{S} = 2G\pedix{M}{BH}/\apix{c}{2}$) .

The quality of the available data prevents a more detailed spectral analysis, but 
the ionization level of the Fe responsible of the absorption and the observed EW$\sim -230\,$eV 
(though with large uncertainties)
would suggest that we are dealing with a high-ionization (ionization parameter\footnote{The ionisation parameter is defined as
$\xi=\pedix{L}{ion}/nR^2$, where $n$ is the hydrogen number density of the gas (in cm$^{-3}$) 
and $R$ is the radial distance of the absorbing/emitting material from the central source of X-ray (in cm), while the ionising luminosity 
\pedix{L}{ion} has units of \lum.} $\log\xi\sim 3\,$\ionpar), high-column density ($\nhsym\sim10^{23}\,$\nh) outflow \citep[e.g.,][]{gofford13}.

\section{Summary}\label{sect:summ}
In this paper, we have reported on 
our $0.3-10\,$keV observation of the Seyfert~1.9 galaxy \src, a member of the local LIRG pair known as CGCG 468-002.
The \swift-XRT data have been analysed jointly with the \swift-BAT spectrum, averaged over $70$ months.
The continuum is well described by an intrinsic power law with photon index $\Gamma\sim 1.8$, absorbed by a column of $\nhsym\sim2.4\times10^{22}\,$\nh.
A reflected component is also observed, with a reflection fraction of $\sim 1.4$, while a weak soft scattered component ($\sim 2$\% of scattering fraction) can account for the observed soft emission. 
The de-absorbed luminosity is $\pedix{L}{2-10\kev}\sim4\times10^{42}\,$\lum.

Our \swift\ monitoring (the first observation of this source extending up to energies $E\sim7-10\,$keV), performed in less than $1\,$month,
provides us a tentative detection ($\sim 2.1\sigma$ significance) 
of an absorption trough at a rest-frame energy of $\sim7.8\,$keV.
When the feature is described by a simple Gaussian absorption line, its properties (e.g., energy and EW)
are consistent with 
an origin in
a material moving with a velocity of $\sim 0.11c$.
To our knowledge, this would be the first detection with \swift\ of a previously unknown high-velocity outflow.

The low statistics of the data and the high energy of the observed residuals,
near the higher boundary of the bandpass of XRT, do not allow us to test
more physically consistent models \citep[e.g., grids of photoionized absorbers generated with the XSTAR  photoionization code;][]{kallman04}.
However, if the detection is confirmed, the observed EW and the derived velocity suggest physical parameters typical of an extremely powerful outflow, 
as observed in only a handful of AGN 
\citep[e.g., the Ultra Luminous Infrared Galaxy/quasar \object{Mrk~231};][]{feruglio15}.
In this case, the kinetic output could 
match or exceed the typical fraction of bolometric luminosities required for AGN feedback.

In fact, \src\ could resemble \object{Mrk~231}: 
the source studied here
is hosted in a star forming merging system and (possibly) shows evidence of a powerful disk wind.
However, these characteristics are combined in \src\ with a lower level of activity, both of accretion and star formation, than observed in \object{Mrk~231}.
This would make \src\ quite unique among the extremely powerful AGN winds studied so far.
Merging systems as the one hosting this source are the objects where we expect to better observe the interplay between star formation and accretion,
since both phenomena can be triggered by galaxy interactions.
Indeed, they are the objects were the coexistence of disk winds and molecular outflows has been found so far
(e.g.,  \object{Mrk~231}, \citealp{feruglio15};  \object{IRAS~F11119+3257}, \citealp{tombesi15}).
In addition, comparing the accretion and star forming properties reported in Sect.~\ref{sect:src}, \src\ seems to be one of the few examples of a source that, 
after a recent episode of star formation, is in a transition phase between star forming-dominated (HII-LIRG) and 
accretion-dominated (Seyfert-LIRG) state \citep{aherrero13}.
The quenching of the star formation can be related (at least partly) to the increasing of the AGN activity, as expected from the co-evolution models.
An outflow powerful enough to affect the environment beyond the SMBH's gravitational sphere of influence, as the one 
possibly detected in the XRT data, could in principle play a significant role in this process.

Better statistics and 
higher resolution observations, extending at energies above $\sim 10\,$keV, are needed in order to 
confirm the presence of the feature,
improve the significance of this detection and
asses the properties of the associated wind, namely the column density and the ionisation state, and then the radial location with respect to the central source.
The knowledge of these parameters would allow us to estimate the mass outflow rate and the kinetic power, to be compared with the 
energetic of the accretion.


\begin{acknowledgements}
We are grateful to the referee for her/his constructive comments that improved the paper.
We warmly thank Alessandro Caccianiga for useful discussions.
We also want to thank Neil Gehrels, Boris Sbarufatti and the \swift\ Mission Operation Center to make every effort to get our observations scheduled.
This work made use of data supplied by the UK \swift\ Science Data Centre at the University of Leicester.
This research has made use of NASA's Astrophysics Data System.
Support from the Italian Space Agency is acknowledged 
(contract ASI INAF I/037/12/0).
The authors acknowledge financial support from the Italian Ministry of Education, Universities and Research (PRIN2010-2011, grant n. 2010NHBSBE).
\end{acknowledgements}

\bibliographystyle{aa} 
\bibliography{ms_swj050819} 

\listofobjects


\end{document}